\documentstyle[12pt,psfig]{article}

\def\eq#1{(\ref{#1})}
\def\order#1{{\cal O}\left(#1\right)}
\def\address #1{\begin{center} {\em #1}\end{center}}
\def\title #1{\begin{center} {\large \bf #1}\end{center}}
\def\author #1{\begin{center} {\large \bf #1}\end{center}}

\def\ba{\begin{eqnarray}}
\def\ea{\end{eqnarray}}
\def\dd{{\rm d}}

\begin{document}
\title{
%
%
\[ \vspace{-2cm} \]
\noindent\hfill\hbox{\rm  } \vskip 1pt
\noindent\hfill\hbox{\rm Alberta Thy 11-01} \vskip 1pt
\vskip 10pt
%
Muonium hyperfine structure and hadronic effects}

\author{Andrzej Czarnecki%
        }
\address{
Department of Physics, University of Alberta\\
Edmonton, AB\ \  T6G 2J1, Canada\\
E-mail: czar@phys.ualberta.ca}

\author{Simon I.~Eidelman}

\address{ Budker Institute for Nuclear Physics,
\\
and Physics Department, Novosibirsk University,
\\
Novosibirsk, 630090, Russia\\
E-mail: simon.eidelman@cern.ch}

\author{Savely G. Karshenboim}

\address{
D. I. Mendeleev Institute for Metrology (VNIIM)\\  
St. Petersburg 198005,
Russia\\
and\\
Max-Planck-Institut f\"ur Quantenoptik,\\
85748 Garching, Germany\\
E-mail: sek@mpq.mpg.de}

\vspace*{12mm}

\begin{abstract}
A new result for the hadronic vacuum polarization correction 
to the muonium hyperfine splitting (HFS) is presented:
$$
   \Delta \nu(\mbox{had -- vp})  = (0.233 \pm 0.003)\, \mbox{kHz}.
$$
Compared with previous calculations, the accuracy is improved
by using the latest data on $e^+e^- \to$ hadrons. The status of the
QED prediction for HFS is discussed.
\end{abstract}


\newpage

\section{Introduction}

Hyperfine structure of two-body atomic systems is of interest for a
variety of reasons.  In particular, it offers opportunities for
precise tests of bound-state QED and accurate determination of
fundamental constants like the fine structure constant $\alpha$,
or electron-to-muon mass ratio.  The hyperfine
interval in the hydrogen ground state used to be the most precisely
measured quantity. The fractional uncertainty was about $10^{-12}$
(see e.g. \cite{ramsey}), although the 
theory cannot make a prediction with
accuracy better than $10^{-5}$.  The theoretical uncertainty comes
from the proton magnetic form factor unknown at low momentum transfers
\cite{karsh99} and from the proton polarizability contribution.

Studying purely leptonic systems such as muonium (the bound state
of a positive muon and an electron), 
one can avoid problems of the
proton structure. Despite the short muon 
lifetime (about 2.2 $\mu$sec), the hyperfine splitting (HFS) of the
muonium ground state has been measured very precisely \cite{Exp,mariam},
\begin{equation}  
\nu_{\rm HFS}({\rm exp}) = 4\,463\,302.776(51)~{\rm kHz},
\label{exp}
\end{equation}
with a fractional uncertainty of 0.011 ppm.

The theoretical prediction for the HFS is determined by the Fermi
energy, resulting from a non-relativistic interaction of electron and
muon magnetic moments. It can be expressed  as a combination of 
fundamental constants
\begin{equation}  \label{EF}
\nu_F=\frac{16}{3}{(Z\alpha)}^2 c Z^2R_{\infty} \frac{m}{M}
\left [ \frac{m_R}{m} \right ]^{3}(1 + a_{\mu} )\,.
\end{equation}
Another convenient formula which will be used to determine the Fermi
energy is 
\begin{equation}  \label{EF1}
\nu_F=\frac{16}{3}{(Z\alpha)}^2 c Z^2R_{\infty} \frac{\mu_\mu}{\mu^e_B}
\left [ \frac{m_R}{m} \right ]^{3}\,.
\end{equation}
The accuracy of these expressions is limited by the electron-to-muon
mass ratio $m/M$ or the ratio of the muon magnetic moment to the
electron Bohr magneton ${\mu_\mu}/{\mu^e_B}$. 
Since the muon anomalous magnetic moment $a_\mu$ is known with 
high accuracy
\cite{BrownMG,CzarneckiPV,kinoshita}, these two formulae are essentially
equivalent. Given that the three most precise experiments
\cite{Exp,mariam,klempt} determined
the magnetic moment ratio, we will use the equation~(\ref{EF1}). 
Using the values $1/\alpha =137.035\,999\,58(52)$ \cite{kinoshita}, 
$cR_\infty =3\,289\,841\,960\,368(25)$ kHz \cite{Mohr99}, ${\mu_\mu}/{\mu_p}
=3.183\,345\,24(37)$ \cite{Exp,mariam}, 
and ${\mu_p}/{\mu^e_B} =1.521\,032\,203(15)\cdot 10^{-3}$ \cite{Mohr99}, 
we find
\begin{equation}  \label{EF2}
\nu_F=4\,459\,031.920(511)(34) \; \mbox{kHz},
\end{equation}
where the first (larger) uncertainty is due to the magnetic moment
ratio and the second to the fine structure constant. 
The value  $M/m = 206.768\,276(24)$ used in this work 
was obtained by us from the values of
the magnetic moments above as well as the experimental value 
$a_{\mu} =1.165\,920\,3(15)\cdot 10^{-3}$ \cite{BrownMG}.
Here $Z$ is the
nuclear charge and in the case of muonium $Z=1$, however, it is
convenient to keep that for the classification of different
contributions (see Section \ref{QEDsec}). In practice, precise
measurements of the muonium HFS \cite{Exp,mariam} serve as the most
precise determination of the muon-to-electron mass ratio
\cite{Mohr99}. In the future, muon mass might be determined more
precisely from other experiments \cite{bosh}, thus enabling a new test
of bound-state QED.

\begin{figure}[htb] 
\hspace*{55mm}
\begin{minipage}{16.cm}
\psfig{figure=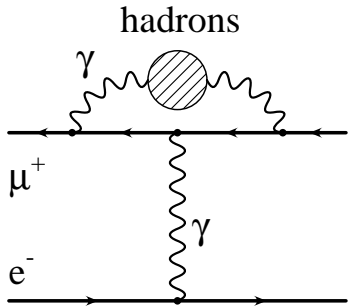,width=45mm}
\end{minipage}
\caption{Hadronic contribution to muonium HFS via the muon
anomalous magnetic moment.}
\label{fig1}
\end{figure}

\begin{figure}[htb] 
\hspace*{32mm}
\begin{minipage}{16.cm}
\begin{tabular}{c@{\hspace*{10mm}}c}
\psfig{figure=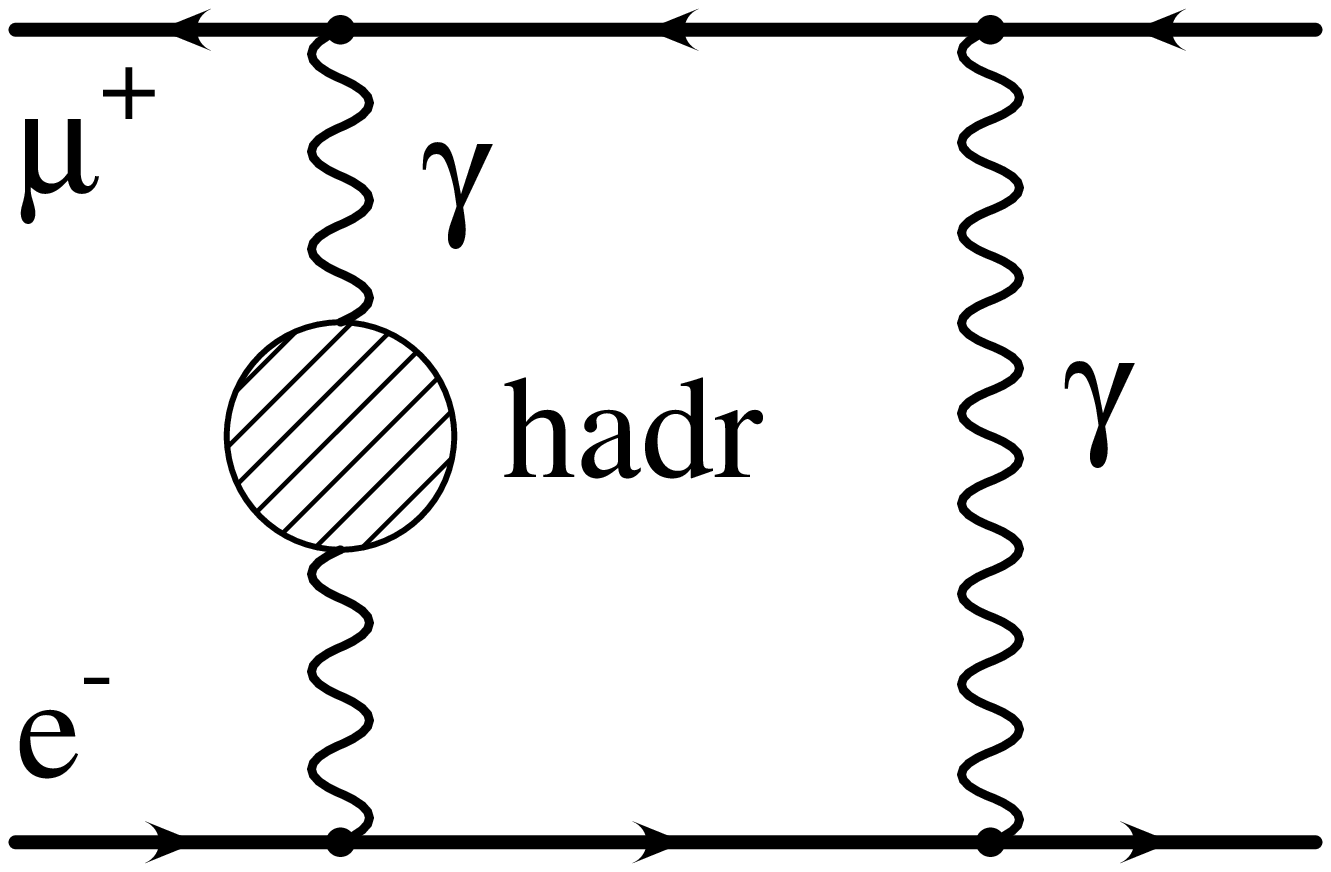,width=40mm}
&
\psfig{figure=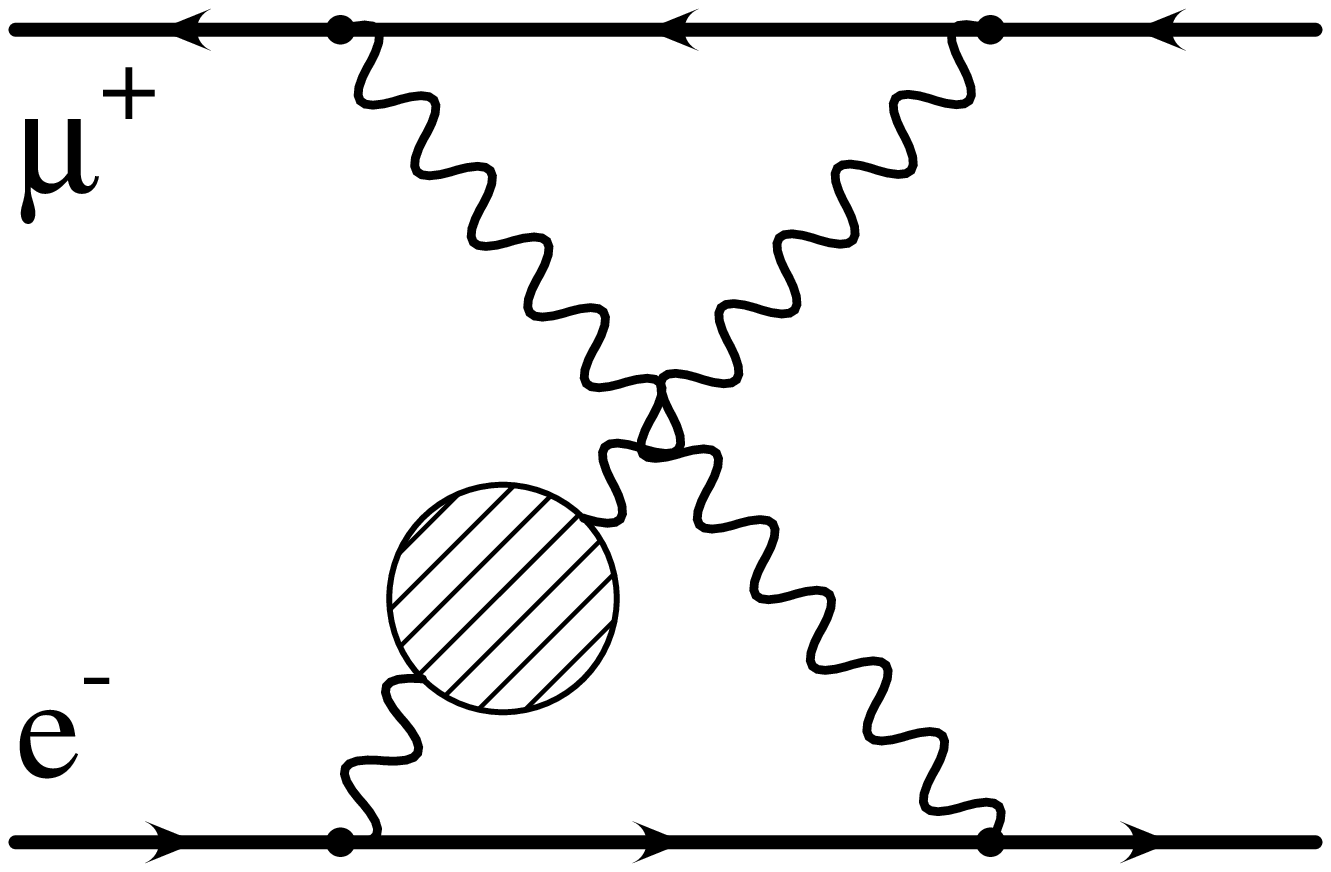,width=40mm}
\end{tabular}
\end{minipage}
\caption{Bound-state hadronic effects to muonium HFS considered
in this paper.}
\label{fig2}
\end{figure}

Further experimental progress in the muonium HFS is also expected
\cite{bosh} if high-intensity muon sources become available.  In view
of those advances, it is conceivable that the accuracy of bound-state
QED tests may become limited by interactions beyond QED, in particular
by the strong interaction effects.  In this paper we study hadronic
vacuum polarization, which is the leading contribution of the strong
interactions.  These effects influence the HFS in two ways: through
effects included in the Fermi energy (contributions to the muon
$a_\mu$ shown in Fig.~\ref{fig1}) and bound-state effects depicted in
Fig.~\ref{fig2}.  Hadronic effects in $a_\mu$ are relatively large
($\sim \alpha^2$) and have been studied by a number of authors over
the last 3 decades (see references in \cite{CzarneckiPV,MarcianoQQ}).
In the present case, we are using an experimental determination of the
muon magnetic moment ($\mu_{\mu}$) and do not address this theoretical 
issue.  We focus on the bound-state effects of the hadronic vacuum 
polarization ($\sim \alpha^2(m/M)$).

The theoretical expression for the hyperfine splitting can be written
in the form
\ba
\nu_{\rm HFS}({\rm theor}) = \nu_F + \Delta\nu({\rm QED})+
\Delta\nu({\rm weak})+ \Delta\nu({\rm had}).
\ea
The biggest correction to the Fermi splitting ($\nu_F$) comes from the
QED effects ($\Delta \nu({\rm QED})$) and it is under consideration in
Section \ref{QEDsec}. The two terms beyond QED are due to weak
($\Delta \nu({\rm weak})$) and strong ($\Delta \nu({\rm had})$) interactions.

Influence of the weak interaction effects on the muonium HFS was
studied in the leading order \cite{weak,kino94}. The leading
correction is induced by the neutral currents, given by a
$Z$-boson exchange diagram.  Recently it was found that the sign of the
corrections had not been well understood and  the absolute value was
verified. In a recent paper \cite{eide} that problem was solved. The
contribution is 
\ba\label{weak} 
\Delta\nu({\rm weak}) = - \frac{G_F m
M}{\sqrt{2}}\frac{3}{4\pi Z\alpha}\cdot\nu_F \simeq - 0.065~{\rm
kHz}\;, 
\ea 
where $G_F$ is the Fermi constant of the weak
interaction. Next--to--leading--order contributions, studied in
Ref.~\cite{kars96}, are $\order{1\%}$ of the leading contribution
(\ref{weak}) and thus negligible.

The strong interaction effects (Fig.~\ref{fig2}) were studied in
\cite{sapi84,kari,fau}.  We examine the hadronic contribution in
Section~\ref{Hardsec} following the approach of Ref.~\cite{KSu}, in
which the QED and hadronic parts are separated and the final result 
in the leading order is presented in the form
\ba\label{hadr}
\Delta \nu(\mbox{had -- vp}) =
 -{ Z\over 2\pi^3}{m\over M}\,\nu_F
\int_{4m_{\pi}^2}^{\infty} \dd s~ \sigma(s)~H(s),
\ea
where $s$ is the center-of-mass 
energy squared, $\sigma(s)$ is the total cross section 
of electron-positron annihilation into hadrons and $H(s)$ is the QED 
kernel calculated in \cite{KSu}. Our calculation gives
\ba
\Delta \nu(\mbox{had -- vp}) = (0.233 \pm 0.003)\, \mbox{ kHz}.
\ea
The uncertainty of this result, based on recent electron-positron
annihilation data, is less than half of that of the most accurate
previous calculation \cite{fau}.  The higher-order hadronic
corrections are considered in part in \cite{KSu}.

In Section~\ref{Sumsec} we discuss our results.

\section{QED contributions \label{QEDsec}}

The problem of the theoretical accuracy of QED effects in the muonium
HFS was considered in \cite{kars96}, where the importance of
higher--order logarithmic corrections was pointed out.  The complete
QED theory includes a variety of contributions: electron and muon
magnetic moments, external field terms, recoil and radiative-recoil
corrections. The theoretical expression can be presented as an
expansion in small parameters: $\alpha$, $Z\alpha$ and $m/M$. In the
case of muonium $Z=1$, however, it is customary to keep $Z$ in 
order to distinguish between 
QED effects (which are counted by powers of $\alpha$)
and the binding Coulomb effect which leads to an appearance of
$Z\alpha$.  In some papers the muon charge is denoted by $Ze$ and so
the QED effects due to the radiative corrections on the muon line
involve $Z^2\alpha$.  Since that is still a QED effect, in our paper
this correction is  just $\alpha$.

Here we discuss briefly the  progress made  since the publication of
Ref.~\cite{kars96}. 
We divide the QED effects in three parts,
\ba
\Delta\nu({\rm QED}) =  \Delta\nu(a_e)+ \Delta\nu({\rm QED3})+
\Delta\nu({\rm QED4}),
\ea
because of their different status. The first term,
\ba
 \Delta \nu(a_e) = a_e \, \nu_F= 5\,170.926(1) \; \mbox{kHz} ,
\label{QEDae}
\ea
arises due to the
electron anomalous magnetic moment $a_e=1\, 159\, 652\, 187 (4) \cdot
10^{-3}$ for which we use the 
experimental value \cite{ae}.  The next term, $\nu({\rm QED3})$
includes the bound-state QED effects up to the 3rd order in the relevant
small parameters, to be discussed in Section \ref{third}.  These
effects are known with good accuracy.  The last term,  $\nu({\rm
QED4})$, is related to the 4th order contributions and is subject to
ongoing investigations.  Its present status will be reviewed in
Section \ref{fourth}. 

\subsection{QED up to third order}
\label{third}

The contributions up to the third order in those parameters give
\begin{eqnarray}  
\label{defth}
\lefteqn{\Delta \nu ({\rm QED3})=}
\nonumber \\
 =&&\hspace*{-5mm} \nu_F
\left\{ \frac{3}{2} (Z \alpha )^2
+  
\alpha (Z \alpha ) \left( \ln 2 - \frac{5}{2} \right)
\right. 
\nonumber \\
&&
\quad
+{\alpha (Z \alpha )^2\over \pi}
\left[ - \frac{2}{3} \ln{1\over (Z\alpha)^2}
\left( \ln{\frac{1}{(Z\alpha)^2}} + 4 \ln 2 - \frac{281}{240} \right)
+ 17.122\,339\ldots \right. 
\nonumber \\
&& \qquad\qquad\qquad\qquad\qquad
\left.
\left. -\frac{8}{15}\ln{2}
+\frac{34}{225}
    \right] \quad 
+0.7718(4)\,\frac{\alpha^2(Z\alpha)}{\pi} 
 \right\}
\nonumber \\
&+& \frac{\nu_F}{1 + a_{\mu}}\frac{(Z\alpha)m}{M} 
\left\{\left[- \frac{3}{\pi}\ln {\frac{M}{m}}
+ (Z\alpha) \left( \ln {\frac{1}{(Z\alpha)^2}}
- 8 \ln 2 + \frac{65}{18} \right) \right]
\right.
\nonumber \\
&&\qquad\qquad \qquad\quad + \left.
\frac{\alpha}{\pi^2}
\left[ - 2 \ln^2 {\frac{M}{m}} + \frac{13}{12} \ln {\frac{M}{m}}
+ \frac{21}{2} \zeta (3) + \frac{\pi^2}{6} + \frac{35}{9}
\right]\right\}\,.
\end{eqnarray}
Origins of individual terms in
Eq.~(\ref{defth}) are discussed in detail in the review \cite{EGS}.
The numerical value resulting from (\ref{defth}) is
\ba
\Delta \nu ({\rm QED3})=-899.557 \; \mbox{kHz},
\label{QED3}
\ea
with an error beyond the displayed digits. 

\subsection{Fourth order}
\label{fourth}

The essential part of the fourth order corrections has been evaluated
(see e.g. \cite{icap}). The known terms are summarized in Table
\ref{Tfoco}.  Some of them were calculated in the leading logarithmic
approximation (using $\ln {1/\alpha}\sim \ln {M/m} \sim 5$) and in
those cases we follow \cite{kars96} and estimate the uncertainty due
to the non-leading terms as half of the leading ones.  Table
\ref{Tfoco} contains a number of recent results which appeared after
the review \cite{kars96}.  Some of the contributions will be reviewed
in more detail below.  We also present in that Table a very small
contribution of the tau lepton loops replacing the hadronic loop in
Fig.~\ref{fig2}.  This effect is $\order{\alpha(Z\alpha){mM/
m_\tau^2}}$ and amounts to about $2\cdot 10^{-3}$ kHz.

\begin{table}[th]
\begin{center}
\begin{tabular}{crr}
\hline
\hline &\\[-1ex]
Correction & Contribution to the HFS & Reference \\[1ex]
\hline &&\\[-1ex]
$(Z\alpha)^4$ &~0.03 kHz &\protect{\cite{brei}}\\[1ex]
$(Z\alpha)^3\frac{m}{M}$ & $-$0.29(13) kHz &
\protect{\cite{kars93,kars93a,kino94,kars96,my2001,hill}} \\[1ex] 
$(Z\alpha)^2\left(\frac{m}{M}\right)^2$ & $-$0.02(1) kHz &
\protect{\cite{lepa77,pach00}} \\[1ex] 
$(Z\alpha)\left(\frac{m}{M}\right)^3$ & $-$0.02 kHz &
\protect{\cite{arno}} \\[1ex] 
$\alpha(Z\alpha)^3$ & $-$0.52(3) kHz & see Table \protect{\ref{Tone}} \\[1ex]
$\alpha(Z\alpha)^2\frac{m}{M}$ & ~0.39(17) kHz &
\protect{\cite{kars93,my2001,hill}} \\[1ex] 
$\alpha(Z\alpha)\left(\frac{m}{M}\right)^2$ & $-$0.04 kHz &
\protect{\cite{eide99,newest}} \\[1ex] 
$\alpha^2(Z\alpha)^2$ & $-$0.04(2) kHz & \protect{\cite{kars93}} \\[1ex]
$\alpha^2(Z\alpha)\frac{m}{M}$ & $-$0.04(3) kHz &
\protect{\cite{eide84,eide89}} \\[1ex] 
$\alpha^3(Z\alpha)$ & $\pm 0.01$ kHz & see caption \\[1ex]
$\tau$ lepton & $ 0.002$ kHz & \protect\cite{sapi84} \\[1ex]
\hline &&\\[-1ex]
Total  & $-0.55(22)$ kHz & \\[1ex]
\hline
\hline
\end{tabular}
\caption{\label{Tfoco} The fourth order corrections.  
The first column gives the order of corrections relative to $\nu_F$.  
The uncertainty due to the
unknown term $\alpha^3(Z \alpha)$ is estimated as
$(\alpha/\pi)^2$ times the known $\alpha(Z \alpha) \nu_F$
term in eq.~\eq{defth}.}
\end{center}
\end{table}

\subsection{One--loop corrections}
\label{Sect23}

The electron one--loop self--energy has been studied in some
detail. After direct calculations of the $\alpha(Z\alpha)^2$ term
\cite{pach96,nio}, two independent exact (without an expansion in
$(Z\alpha)$) numerical evaluations were performed
\cite{blun,ys2001}. Unfortunately, the accuracy for $Z=1$ was not high
in both cases and the final results for muonium were obtained by
extrapolating
the data from higher values of $Z$ to $Z=1$, using previously found
coefficients of $\alpha(Z\alpha)^2$ \cite{pach96,nio} and
$\alpha(Z\alpha)^3\ln{Z\alpha}$ \cite{kars96}. The result of the later
calculation \cite{ys2001} is presented in  Table \ref{Tone} as
$\alpha(Z\alpha)^{3+}$. It contains the non-logarithmic
$\alpha(Z\alpha)^{3}$ term and higher-order corrections.
The result of \cite{blun} is slightly higher than that of \cite{ys2001}.
The calculation gives for this term
$-14.3(1.1)\,\alpha(Z\alpha)^3\nu_F/\pi$ \cite{ys2001} and
$-12(2)\,\alpha(Z\alpha)^3\nu_F/\pi$ \cite{blun}. Recently the
non-logaricthmic term was
calculated directly by expansion in $Z\alpha$ to be
$-15.9(1.6)\alpha(Z\alpha)^3\nu_F/\pi$ \cite{Nio2001}.

The uncertainty of the extrapolation
\cite{ys2001} is determined as an estimate of the 
unknown higher-order terms rather than 
the accuracy of numerical data. However, the final uncertainty is
not clear, because of the value of the non-logarithmic
$\alpha(Z\alpha)^2$ term \cite{pach96,nio}
(cf.~\cite{sapi83a,kars96}):
\ba
\bigl(17.122\,339\dots\bigr)\cdot\frac{\alpha(Z\alpha)^2}{\pi}\,\nu_F \quad
\cite{pach96}\,, 
\nonumber \\
\bigl(17.122\,7\pm0.001\,1\bigr)\cdot\frac{\alpha(Z\alpha)^2}{\pi}\,\nu_F 
\quad \cite{nio}\,.
\ea
These results appear to be in fair agreement with each other.
However, it has been pointed out \cite{nio} that the term-by-term
comparison reveals inconsistencies. The low--energy contribution in
Ref.~\cite{nio} is by 0.0257(2) higher  than that in
Ref.~\cite{pach96}, while the medium-- and high--energy parts are
by 0.0246(14) lower.  It should also be mentioned that different fits
for extrapolation to the zero photon mass $\lambda$ in Ref.~\cite{nio}
are not consistent. Such extrapolation was necessary to
calculate the high momentum part. These problems remain to be
clarified. Since the same method was used in \cite{Nio2001},
it is the result of \cite{ys2001} that we include to Table \ref{Tone}. 

\begin{table}[th]
\begin{center}
\begin{tabular}{cccc}
\hline\hline &&&\\[-1ex]
Order & Contribution to the HFS & Reference & Comments \\[1ex]
\hline &&&\\[-1ex]
$\alpha(Z\alpha)^3\,\ln{\frac{1}{Z\alpha}}$
& $-$0.53 kHz  & \protect{\cite{kars96}} & self--energy\\[1ex]
$\alpha(Z\alpha)^{3+}$
&   $-0.06$ kHz & \protect{\cite{ys2001}} & self--energy\\[1ex]
$\alpha(Z\alpha)^3\,\ln{\frac{1}{Z\alpha}}$
& $\phantom{-}0.03$ kHz & \protect{\cite{kars96}} & vacuum polarization \\[1ex]
$\alpha(Z\alpha)^3$
&  $\phantom{-}0.03$ kHz & \protect{\cite{kars99}} & vacuum
polarization \\[1ex] 
$ \alpha(Z\alpha)^3$
& $\pm 0.03$ kHz &  see Sect.~\protect\ref{Sect23}  & Wichmann--Kroll\\[1ex]
\hline &&&\\[-1ex]
one-loop & $-0.52(3)$ kHz  && sum of all terms\\[1ex]
\hline\hline
\end{tabular}
\caption{\label{Tone} Higher--order one--loop contributions. 
}
\end{center}
\end{table}

Recently, the one--loop vacuum polarization was calculated exactly (to
all orders in $Z\alpha$) \cite{kars98,kars99}. For our purpose it is
enough to know the $\alpha(Z\alpha)^3$ terms only
\begin{equation} \label{vp13}
\Delta \nu = \alpha(Z\alpha)^3
\left(\frac{13}{24}\ln\frac{2}{Z\alpha}+\frac{539}{288}\right)
\nu_{F}
\,,
\end{equation}
the logarithmic part of which was obtained in Ref.~\cite{kars96}, while the
constant was found in Ref.~\cite{kars99}.

The only unknown term in the order $\alpha(Z\alpha)^3$ is now the
so called Wichmann--Kroll contribution. We estimate it by the value of the
non-logarithmic part of the VP term in Eq.~(\ref{vp13}).

\subsection{Recoil effects}

A number of results on various recoil effects were obtained after the
publication of \cite{kars96}. 
\begin{itemize}
\item
Pure recoil effects of the order $(Z\alpha)^2$ were studied without
an $m/M$ expansion in Ref.~\cite{pach97} for arbitrary mass and numerical
results for several values of $m/M$ were obtained. Recently the same
correction was found for muonium \cite{pach00} (see Table \ref{Tfoco}).

\item
For radiative recoil corrections ($\alpha(Z\alpha)m/M$), there are
minor discrepancies between the published numerical and analytical
results, summarized in Table \ref{Trrc}.  All analytical results were
obtained by two groups which are in agreement.  It is likely that the
uncertainty of the numerical calculations was underestimated
\cite{sapi99} and the discrepancies have no connection with
higher-order corrections.  This explanation is supported by the
similar situation with the radiative recoil corrections in
positronium, shown in the same table.

\begin{table}[th]
\begin{center}
\begin{tabular}{cccc}
\hline \hline &&&\\[-1ex]
Contribution & Numerical & Analytical & Discrepancy \\[1ex]
             & result    & result   &              \\[1ex]
\hline &&&\\[-1ex]
e$-$line & ~3.335(58)  \protect{\cite{sapi83,sapi84}} & ~3.499
\protect{\cite{eide86,broo,eide99,newest}}& 0.164(58)\\[1ex] 
$\mu-$line & $-$1.0372(91)  \protect{\cite{sapi83,sapi84}} & $-$1.0442
\protect{\cite{eide88,eide99}} 
&$-0.0070(91)$\\[1ex]
Positronium &  $-1.787(4)$ \protect{\cite{sapi83,sapi84}} &  $-1.805$
\protect{\cite{pach98,Czarnecki:1998zv,Czarnecki:1999mw}} 
& $-$0.018(4)  \\[1ex]
\hline\hline
\end{tabular}
\caption{\label{Trrc} Radiative--recoil corrections in units of
$\alpha^2(m/M)\nu_F$.  The numerical results include all orders in
$m/M$.  The analytical results refer to sums of the third and fourth
order, $\alpha^2(m/M)$ and $\alpha^2(m/M)^2$ for muonium.  
For positronium, $M=m$ and $\nu_F$ is defined in (\protect\ref{EF}).}
\end{center}
\end{table}
\item 
The uncertainty of the QED calculations of the muonium hyperfine
structure arises mainly from two sources: $\alpha(Z\alpha)^2m/M$ and
$(Z\alpha)^3m/M$. Only double logarithmic corrections were calculated
until recently, when some single-logarithmic terms ($\sim
\ln(Z\alpha)$) were found. The single-logarithmic terms with a recoil
logarithm ($\ln(M/m)$) are still unknown.  We discuss a possible
consequence for the accuracy of the theoretical calculations in the
next section.
\end{itemize}

\subsection{Non-leading logarithmic terms and uncertainty of the
calculations} 

In this paper we follow \cite{kars96} and estimate any unknown
non--leading terms by a half--value of the corresponding leading
logarithmic terms and sum them as independent. There is no proof that
the non-leading terms can be estimated in this way.  However, all
experience with different contributions to the Lamb shift and the
hyperfine structure (see e.g.~the review \cite{EGS} with a collection of
various terms) demonstrates that this assumption is reasonable.

The fourth order leading corrections contain cubic and quadratic
logarithmic terms and in the case of some contributions, known in the
logarithmic approximation, more than the leading term has been
evaluated (see Table \ref{Tfoco} for references):
\[
\frac{\alpha^2(Z\alpha)}{\pi^3}\frac{m}{M}\nu_F\times
\left\{-\frac{4}{3}\,\ln^3{\frac{M}{m}} 
+\frac{4}{3}\,\ln^2{\frac{M}{m}}\right\}\;,
\]
\[
\frac{\alpha(Z\alpha)^2}{\pi}\frac{m}{M}\nu_F\times\left\{
\frac{16}{3}\,\ln^2{{\frac{1}{Z\alpha}}}
+ \left(1+\frac{32}{3}\ln2 - \frac{431}{90}
\right)\,\ln{\frac{1}{Z\alpha}}\right\}
\]
and
\[
\frac{(Z\alpha)^3}{\pi}\frac{m}{M}\nu_F\times\left\{
-3\,\ln{\frac{M}{m}}\,\ln{\frac{1}{Z\alpha}}
-\frac{2}{3}\,\ln^2{\frac{1}{Z\alpha}}
+
\left(\frac{101}{9}-20\ln2\right)\,\ln{\frac{1}{Z\alpha}}\right\}
\,.
\] 
Even in this case we estimate the unknown terms as the half-value of the
{\em leading} term. From comparison with e.g.  known contributions of
the third order $\alpha(Z \alpha)^2\nu_F$ (see Eq. \eq{defth}) one
sees that the constant term can be even bigger than the non-leading
logarithm.  Error estimates based on a non-leading logarithmic term
may be misleading. The reason is that the leading logarithm
in most cases originates from relatively
simple diagrams under well understood conditions. The cancelations between
different contributions occur very seldom. The leading
logarithmic term has a {\em natural} value and is useful for
estimates. In case of a cancelation which, although rare, 
is possible, an estimate should be based on the
contributions {\em before} the cancelation. In case of the
non-leading logarithmic terms, there are usually a few sources of
different nature and some cancelations take place very often. 

The total magnitude of the muonium HFS interval calculated within QED 
\ba
   \nu_{\rm QED} = 4\,463\,302.738(511)(34)(220)\, \mbox{ kHz}.  
\label{QED}
\ea
is found by adding the
values given by Eqs.~\eq{QEDae}, \eq{QED3}, and the sum of
the 4th order contributions listed in Table \ref{Tfoco} to the Fermi
splitting \eq{EF2}.
The third uncertainty is due to the 4th order QED effects  in Table
\ref{Tfoco}.

\section{Hadronic contributions \label{Hardsec}}

\subsection{Calculation of the hadronic contributions to muonium HFS}

The lowest order hadronic contribution to the muonium hyperfine 
splitting is given by the following expression:
\ba
\Delta \nu(\mbox{had -- vp}) =
 -{1 \over 2\pi^3}{m\over M}\,\nu_F
\int_{4m^2_{\pi}}^{\infty} \dd s~ \sigma(s)~H(s),
\label{eq:hs}
\ea
where the kernel $H(s)$ calculated in \cite{KSu} is:
\ba 
H(s) &=& \left( {s\over 4M^2}+2\right) r \ln{1+r\over 1-r}
- \left( {s\over 4M^2}+{3\over 2}\right)\ln {s\over M^2}+{1\over 2},
\qquad r\equiv \sqrt{1-{4M^2\over s}}.
\label{eq:hadMu}
\ea
The bulk of those effects is computed using the experimental data on
the cross sections of $e^+e^- \to {\rm hadrons}$ in the energy range
$\sqrt{s}<\sqrt{s_0}$ where $s_0$ is a scale above which
perturbative formulae can be used.  We choose $\sqrt{s_0}=12$ GeV. After
that one performs direct numerical integration of the experimental points
similar to the approach of \cite{ej} where hadronic corrections
to $a_{\mu}$ were calculated.  In contrast to
the methods in which some approximation of the data is used for the
integration, in our approach model dependence is avoided as much as 
possible and, moreover, the calculation of the uncertainties is 
straightforward.  In addition to the data set used in \cite{ej}, 
one can take into account significant progress in the
measurement of the hadronic cross sections in the energy range below 1.4 GeV
with two detectors at VEPP-2M in Novosibirsk \cite{akh,ser}
and the $R$ determination in the energy range 2 to 5 GeV by the BES
detector in Beijing \cite{bes}.
   
The integration procedure gives for the contribution of this part  
\ba
\Delta \nu({\rm bulk}) = (0.2031 \pm 0.0031)\; {\rm kHz}.
\label{hfs:bulk}
\ea

The narrow resonances ($\omega, \phi, J/\psi$- and
$\Upsilon$-families) are evaluated separately. In a zero width
approximation the contribution of a resonance B with mass $m_B$ and
electronic width $\Gamma_{ee}$ is
\ba
\Delta \nu(\mbox{resonance B}) & =&  
-{6\over \pi} 
{m\over M} \,\nu_F\, {\Gamma(B\to e^+e^-)  \over m_B}  H(m_B^2) \cdot \left
[ 1-\Delta \alpha(m_B^2)\right]^2 ,
\nonumber \\
\Delta \alpha(s) &=& {\alpha\over 3\pi} \sum_f Q_f^2 N_{cf} \left(
\ln{s\over m_f^2} -{5\over 3} \right).
\ea

The leptonic widths relevant for our study should correspond to the lowest
order (Born) graphs. However, experimentally measured leptonic widths
listed and averaged in the Review of Particle Physics \cite{pdg}, 
contain an additional contribution of the vacuum polarization by leptons
and hadrons. Therefore, for the transition to the lowest order widths 
one should multiply the experimental values by
$\left[1-\Delta \alpha(m_B^2)\right]^2$  (see the discussion of this
issue in \cite{ej}). $Q_f$ and
$m_f$ are the charge and mass of the fermion $f$, and $N_{cf}$ is the
number of colours for the corresponding fermion ($N_{cf}=1$ for
leptons).  The above approximate formula for $\Delta \alpha(s)$ is
valid for $s\ll M_W$ and describes contributions of fermions much
lighter than $\sqrt{s/4}$.  In the present calculation we consider
only effects of the lightest leptons, $e$ and $\mu$, and neglect the
hadronic contributions to $\Delta \alpha(s)$.  Hadronic loops are
relevant only for heavy flavours (the $J/\psi$- and $\Upsilon$-family
of resonances), which however give a smaller contribution and have a
larger relative error.  The hadronic effects, had they been included,
would have shifted the $J/\psi$-family contribution by a few
percent. The $\Upsilon$-family contributes  about $2 \cdot
10^{-3}$ of the total resonance contribution.

The sum of the individual resonance contributions gives
\ba
\Delta \nu(\rm {res}) = 
(0.0290 \pm 0.0006) \; {\rm kHz}.
\label{hfs:reson}
\ea

In the region above $s_0$ one can use a perturbative formula for the 
hadronic cross section
\ba
\sigma(s) &=& {4\pi\alpha^2\over 3s} \cdot R(s),
\nonumber \\
R(s) &=& R^{(0)}(s) \left[1+ \frac {\alpha_s} {\pi} + C_2 
\left(\frac {\alpha_s} {\pi}\right)^2 +
C_3 \left(\frac {\alpha_s} {\pi}\right)^3 \right],
\nonumber \\
R^{(0)}(s) &=& N_c\left({1\over 9} N_{1/3} + {4\over 9} N_{2/3} \right)
 = {11\over 3} \quad (s> s_0),
\ea
where $N_{1/3(2/3)}$ is the number of ``active'' quarks 
with a charge  1/3 (2/3) at given $s$, $C_2 = 1.411$ and $C_3 = -12.8$
\cite{kat}.    

At  $s\gg M^2$ one can use the asymptotic formula for $H(s)$ \cite{KSu}:
\ba
H(s)\to -{M^2\over s}\left({9 \over 2}\;{\rm ln}{s \over M^2}+{15 \over 4}\right)
\qquad \mbox{(for $s\to \infty$)},
\ea
so that after the numerical integration one  finally obtains
\begin{eqnarray}
\Delta \nu(\rm {cont}) \simeq  0.0012 \mbox{ kHz}.
\label{hfs:cont}
\end{eqnarray}

The final result for the lowest order hadronic contribution is found
as a sum of eqs.~\eq{hfs:bulk},\eq{hfs:reson},\eq{hfs:cont}
\ba
\Delta \nu(\mbox{had -- vp})\!\!\!\!&=& \!\!\!\!
\Delta \nu(\rm bulk)
+\Delta \nu(\rm res)
+\Delta \nu(\rm cont)
=
(0.233\pm 0.003)\, \mbox{ kHz}.
\ea

The individual contributions are presented in
Table \ref{contri}.  It is clear that the
dominant contribution (about 84\%) comes from the very low energy 
range below 1.4 GeV studied in Novosibirsk. The regions below 1.4 GeV
and from 1.4 to 3.1 GeV give about the same contribution to the total
uncertainty. For most of the hadronic channels below 1.4 GeV 
analysis is still in progress in Novosibirsk. However,  
one can hardly expect significant improvement in this region since 
the dominant fraction of the error comes from the 2$\pi$ channel which 
cross section is already known with a very high accuracy 
of about 0.6\% around the $\rho$ meson \cite{logash}. 
Very promising for decreasing the uncertainty of the  2$\pi$ channel
could be use of the $\tau$ lepton data \cite{adh}, but the real
accuracy of this approach has been recently questioned \cite{eid,kirt}.
 The improvement of the
uncertainty for $\sqrt{s}$ from 1.4 to 3.1 GeV will require new
$e^+e^-$ colliders covering this energy range \cite{vena}. Some progress 
in this region is also 
possible due to the  B-factories and CESR which can contribute 
by using hadronic events from initial state radiation \cite{bena,sol}.   

\begin{table}[htb]
\begin{center}
\begin{tabular}{lcc}
\hline \hline &&\\[-1ex]
Final state & Energy range, GeV & $\Delta\nu$, Hz \\[1ex]
\hline &&\\[-1ex]
2$\pi$ & 0.28--1.4 & $158.8 \pm 1.9$ \\[1ex]
$\omega$ &  & $12.4 \pm 0.4$ \\[1ex]
$\phi$ & & $13.2 \pm 0.4$ \\[1ex]
Hadrons & 0.6--1.4 & $10.7 \pm 0.8$ \\[1ex]
Hadrons & 1.4--3.1 & $23.8 \pm 2.2$ \\[1ex]
$J/\psi$ & & $3.4 \pm 0.2$ \\[1ex]
Hadrons & 3.1--12.0 & $9.8 \pm 0.5$ \\[1ex]
Hadrons & $> 12.0$ & 1.2 \\[1ex]
\hline &&\\[-1ex]
Total & & $233.3 \pm 3.1$ \\[1ex]
\hline\hline
\end{tabular}
\caption{\label{contri} Contributions to muonium HFS}
\end{center}
\end{table}

\subsection{Comparison with other calculations}

It is instructive to compare the relative contributions of different
energy ranges for $\nu_{\rm HFS}$ and $a_{\mu}$.  From Table
\ref{contri1} one can see that they are very close to each other
indicating the importance of the low energy regions. This is a natural
consequence of the similar kernel structure, so  that the main contribution 
in both cases comes from the low range of $\sqrt{s}$.  One can also note
some enhancement of the high energy  contribution to the HFS compared to
that to the anomalous magnetic moment. That is due to the different
asymptotic behaviour of the QED kernels for these two problems. 
For $a_{\mu}$ the asymptotics of the kernel is proportional to $M^2/s$, 
whereas for the muonium HFS it contains an additional
logarithmic enhancement.

\begin{table}[htb]
\begin{center}
\begin{tabular}{lccc}
\hline \hline &&&\\[-1ex]
Final state & Energy range, GeV & $\Delta\nu$, \% & $a_{\mu}$, \% \\[1ex]
\hline &&&\\[-1ex]
2$\pi$ & 0.28--1.4 &  68.0 & 71.8 \\[1ex]
$\omega$ &  &           5.3 & 5.7 \\[1ex]
$\phi$ & &          5.7 & 5.8 \\[1ex]
Hadrons & 0.6--1.4 &     4.6 & 4.2 \\[1ex]
Hadrons & 1.4--3.1 &     10.2 & 8.2 \\[1ex]
$J/\psi$ & &          1.5 & 1.3 \\[1ex]
Hadrons & 3.1--12.0 &     4.2 & 2.8 \\[1ex]
Hadrons & $> 12.0$ &   0.5 & 0.2 \\[1ex]
\hline &&&\\[-1ex]
Total & & 100.0 & 100.0  \\[1ex]
\hline\hline
\end{tabular}
\caption{\label{contri1} Contributions to muonium HFS and $a_{\mu}$ }
\end{center}
\end{table}

In Table \ref{compa} we compare results of various calculations 
of the leading order hadronic contribution to the muonium HFS. 
The first estimate of this effect was performed in \cite{sapi84}. 
The authors took into account 
the contributions of the $\rho$, $\omega$ and $\phi$ mesons and 
parameterized the hadron continuum above 1 GeV under  the assumption 
that $R$ is constant and equals 2. This
simplified approach gave  nevertheless a result fairly close
to those of the later, more sophisticated analyses. It is not surprising 
since the dominant part of the hadronic contribution comes just from the
lowest vector mesons which properties were known quite accurately at
that time. In \cite{kari} additional experimental data were taken into account
above 1 GeV, including effects of heavy quarkonia.  The hadron  
continuum was parameterized with a function $R(s)~=~A s^B$ in five 
energy ranges below 47 GeV and above this energy the asymptotic QCD formula 
with six quarks was used. Finally, in the recent work \cite{fau} the authors 
used a similar approach with a slightly more sophisticated parameterization
of the hadron continuum in seven energy ranges below 60 GeV. Unfortunately,
the authors of the above mentioned papers ignore
the model dependence of their results which can be fairly strong and
the error of their calculations only reflects the quality of the 
fit in their rather artificial models describing the data. Even larger is the
effect of the systematic uncertainties completely ignored
in most of the calculations.

\begin{table}[htb]
\begin{center}
\begin{tabular}{lcc}
\hline \hline &&\\[-1ex]
Source & Reference & $\Delta \nu(\mbox{had -- vp})$, kHz \\[1ex]
\hline &&\\[-1ex]
J. R. Sapirstein {\it et al.}, 1984 & \cite{sapi84} & $0.22 \pm 0.03$ \\[1ex]
A. Karimkhodzaev and R. N. Faustov, 1991 & \cite{kari} & $0.250 \pm 0.016$ \\[1ex]
R. N. Faustov {\it et al.}, 1999 & \cite{fau} & $0.240 \pm 0.007$ \\[1ex]
A. Czarnecki {\it et al.}, 2001 & {\em This work} & $0.233 \pm 0.003$ \\[1ex]
\hline\hline
\end{tabular}
\caption{\label{compa} Comparison of various calculations}
\end{center}
\end{table}
 
This is particularly true for the dominant contribution to the muonium
HFS coming from the 2$\pi$ channel. For example, in \cite{fau} its
accuracy calculated within the model of \cite{pich} is unrealistically
high and reaches 0.65\% although the model itself is based on the data
of \cite{barkov} where the systematic uncertainty in the dominant
$\rho$ meson region varies from 2 to 4.4\%. Direct integration of the
experimental points in our work also assumes some model for the energy
dependence of the data. We estimated the model dependence by comparing
the trapezoidal integration in which experimental points are connected
with straight lines with even simpler rectangular integration which
assumes a constant cross section within a small energy range and found
that both methods led to the same result, the difference being much
smaller than the error. Both models allow a relatively simple estimation
of the uncertainty arising because of the systematic errors. As
already noted above, the effect is dominated by the contributions of
the low lying vector mesons. Therefore, in all described calculations
the central values of $\Delta \nu(\mbox{had -- vp})$ are close to each
other. However, in our opinion their uncertainty might have been
underestimated.  The decrease of the uncertainty in the result
presented in this paper became possible thanks to the 
utilization of the most recent data set coming from 
the high precision  $e^+e^-$ experiments in the low energy region.

\section{Conclusions \label{Sumsec}}

\begin{table}[th]
\begin{center}
\begin{tabular}{lr}
\hline
\hline &\\[-1ex]
Correction & Contribution to the HFS  \\[1ex]
\hline &\\[-1ex]
$\nu_F$ &  4\,459\,031.920(511)(34)~kHz  \\[1ex]
$\Delta\nu(a_e)$ &   5\,170.926(1)~kHz\\[1ex]
$\Delta\nu({\rm QED3})$ &  $-899.557$~kHz  \\[1ex]
$\Delta\nu({\rm QED4})$ &     $-0.55(22)$~kHz  \\[1ex]
$\Delta\nu({\rm weak})$ &    $-0.065$~kHz  \\[1ex]
$\Delta\nu({\mbox{had -- vp}})$ &  0.233(3)~kHz  \\[1ex]
$\Delta\nu({\mbox{had -- h.o.}})$ &  0.007(2)~kHz  \\[1ex]
\hline&\\[-1ex]
$\nu_{\rm HFS}({\rm theor})$ & 4\,463\,302{.}913(511)(34)(220)~kHz  \\[1ex]
\hline&\\[-1ex]
$\nu_{\rm HFS}({\rm exp})$ &  4\,463\,302.776(51)~kHz  \\[1ex]
\hline
\hline
\end{tabular}
\caption{\label{Tfinal} The muonium hyperfine structure: various
theoretical contributions  discussed in this paper}
\end{center}
\end{table}

The calculation of the hyperfine splitting in the ground state of
muonium is summarized in Table \ref{Tfinal}. The biggest uncertainty
of the theoretical expression comes from the inaccuracy of
experimental determination of the muon mass and its magnetic
moment. This inaccuracy is essentially related to the statistical
uncertainty of the present experiments \cite{Exp,mariam,klempt}.
Improvement of the statistical accuracy in muonium spectroscopy
studies will be possible with future intense muon sources.
Construction of such facilities is being considered in connection with
muon storage rings which would serve as neutrino factories
\cite{GeerIZ,AutinCI}.  There are plans to build a high-intensity muon
source at the Japanese Hadron Facility \cite{KunoWN}.  The muon
production rates at those future sources are expected to be around
$10^{11}-10^{12} \ \mu/s$, higher than the most intense present beams
(at the Paul Scherrer Institute) by a factor of 300--3000 or more.  If
muonium spectroscopy becomes part of the research program at these
facilities, the statistical errors, which limit the present accuracy,
will be decreased by one--two orders of magnitude \cite{bosh}.  In the
more remote future, if a muon collider is constructed, further
increase of beam intensity by a factor of about 100 will be possible.

In the present paper we have studied the limits of possible
bound-state QED tests with muonium hyperfine structure.  Given the
recent progress in multi-loop QED calculations \cite{EGS,kniga}, one
can expect that higher-order radiative effects will be evaluated when
warranted by the experimental precision.  However, the ultimate
accuracy of every theoretical prediction will be limited by the
knowledge of the hadronic effects.  Here, we have studied the leading
order vacuum polarization effects and found that they can be
calculated with 1.3\% (or 3.1 Hz) accuracy.  Higher order corrections can
modify this contribution by a few percent.  In the leading logarithmic
approximation, they were found to increase the lowest order hadronic
contribution by about 3\% \cite{KSu}.  In the case of the muon
anomalous magnetic moment, this logarithmic correction is not dominant
\cite{kinoshita85}.  At present, it is not clear how good this analogy
is and how reliable the logarithmic approximation is in the case of
the muonium hyperfine splitting.  When the experimental accuracy is
improved, the higher order hadronic effects will have to be
scrutinized.  The evaluation of the diagrams with the vacuum
polarization is straightforward, since only moderate relative accuracy
is sufficient.  The theoretical accuracy will be limited by the
knowledge of the hadronic vacuum polarization in the leading order
diagram, and by the ability to compute the hadronic light-by-light
scattering diagrams which at present is possible only within models of
low-energy hadronic interactions.

\subsection*{Acknowledgements}

We are grateful to M. I. Eides and V. A. Shelyuto for stimulating
discussions.  This work was supported in part by the Natural Sciences
and Engineering Research Council of Canada and by the Deutscher
Akademischer Austauschdienst (A.C.), and by the Russian Foundation
for Basic Researches (under grant \# 00-02-16718) and the Russian
State Program ``Fundamental Metrology'' (S.G.K.). S.I.E. acknowledges
the hospitality of the University of Pittsburgh where part of this work
has been performed.


\begin{thebibliography}{999}

\frenchspacing

\bibitem{ramsey} N. Ramsey, in {\em Quantum Electrodynamics} (Ed. by T.
Kinoshita,
World Sci., Singapore, 1990), pp. 673--696; Hyp. Interactions {\bf 81}, 97 (1993). 

\bibitem{karsh99} S. G. Karshenboim, Can. J. Phys. {\bf 77}, 241 (1999).

\bibitem{Exp} W. Liu {\it et al.},  
Phys. Rev. Lett. {\bf 82}, 711 (1999).

\bibitem{mariam} F. G. Mariam {\it et al.}, 
Phys. Rev. Lett. {\bf 49}, 993 (1982).

\bibitem{BrownMG} 
H.~N.~Brown {\it et al.} (Muon g-2 Collaboration),
Phys. Rev. Lett. {\bf 86}, 2227 (2001).

\bibitem{CzarneckiPV}
A.~Czarnecki and W.~J.~Marciano,
Phys.\ Rev.\ D {\bf 64}, 013014 (2001).


\bibitem{kinoshita} V. W. Hughes and T. Kinoshita,
Rev. Mod. Phys. {\bf 71}, S133 (1999);\\ 
T. Kinoshita, in {\em Hydrogen atom: Precision physics of simple 
atomic systems.} 
Ed. by S. G. Karshenboim et al., (Springer, Berlin, Heidelberg, 2001), 
p.~157.

\bibitem{klempt} E. Klempt {\it et al.},  
Phys. Rev. D {\bf 25}, 652 (1982).

\bibitem{Mohr99}
P.~J. Mohr and B.~N. Taylor, Rev. Mod. Phys. {\bf 72}, 351 (2000).

\bibitem{bosh} M. Boshier, V. Hughes, K. Jungmann, and G. Z. Putlitz, 
Comm. At. Mol. Phys. {\bf 33}, 17 (1996);\\  
K. Jungmann in {\em Hydrogen atom: Precision physics of simple atomic 
systems.} 
Ed. by S. G. Karshenboim et al., (Springer, Berlin, Heidelberg, 2001), 
pp. 81--102.  

\bibitem{MarcianoQQ}
W.~J.~Marciano and B.~L.~Roberts, hep-ph/0105056, 2001.
 
\bibitem{weak} M. A. B. B\'{e}g and G. Feinberg, Phys. Rev. Lett. {\bf 33},
606, (1974); {\bf 35}, 130 (E) (1975);\\
G. T. Bodwin and D. R. Yennie, Phys. Rep. {\bf 43}, 267 (1978).

\bibitem{kino94} T. Kinoshita and M. Nio, Phys. Rev. Lett. {\bf 72}, 3803 
(1994) ; Phys. Rev. D {\bf 53}, 4909 (1996).

\bibitem{eide} M. I. Eides, Phys. Rev. A {\bf 53}, 2953 (1996).

\bibitem{kars96} S. G. Karshenboim, Z. Phys. D {\bf 36}, 11 (1996).

\bibitem{sapi84}  J. R. Sapirstein, E. A. Terray, and D. R. Yennie, Phys. Rev.
D {\bf 29}, 2290 (1984). 

\bibitem{kari} A. Karimkhodzhaev and R. N. Faustov, Yad. Fiz. {\bf 53}, 
1012 (1991) (in Russian) (1991); Sov. J. Nucl. Phys. {\bf 53}, 626 (1991).

\bibitem{fau}
R. N. Faustov, A. Karimkhodzhaev and A. P.~Martynenko, 
Phys. Rev. A {\bf 59}, 2498 (1999).

\bibitem{KSu} S. G. Karshenboim and V. A. Shelyuto, to be published. 
E-print: hep-ph/0107328.
The result for the kernel $H(s)$ was obtained in 1995, but was not 
published at that time. It was quoted in \cite{kars96}.

\bibitem{ae} R. S. van Dyck Jr., P. B. Schwinberg, and H. G. Dehmelt,
Phys. Rev. Lett.  {\bf 59}, 26 (1987).

\bibitem{EGS} M. I.  Eides,  H. Grotch and V. A.  Shelyuto,
Phys. Rep. {\bf 342}, 63 (2001).

\bibitem{icap}S. G. Karshenboim, in {\em Atomic Physics} 
{\bf 17} (AIP conference proceedings {\bf 551}) 
Ed. by E. Arimondo et al. AIP, 2001, pp. 238--253. 
E-print: hep-ph/0007278.

\bibitem{brei} G. Breit, Phys. Rev. {\bf 35}, 1477 (1930).

\bibitem{kars93} S. G. Karshenboim, ZhETF {\bf 103}, 1105 (in
Russian) (1993); JETP {\bf 76}, 541 (1993).

\bibitem{kars93a} S. G. Karshenboim, {\em 25$^{th}$ E.G.A.S.  Conference.
Abstracts.} Caen, 1993.  P1-010; {\em 1994 Conference on precision
electromagnetic measurements. Digest}.  Boulder, 1994. 225.

\bibitem{my2001} K. Melnikov and A. Yelkhovsky, Phys. Rev. Lett. {\bf 86},
1498 (2001).

\bibitem{hill} R. Hill, Phys. Rev. Lett. {\bf 86}, 3280 (2001).

\bibitem{lepa77} G. P. Lepage, Phys. Rev. A {\bf 16}, 863 (1977);\\
G. T. Bodwin, D. R. Yennie and M. A. Gregorio, Phys. Rev.
Lett. {\bf 41}, 1088 (1978).

\bibitem{pach00} K.~Pachucki, unpublished. Quoted as presented  at
{\em International Symposium on Lepton Moments} 
(IWH Heidelberg, June 8-12, 1999)
(http://www.physi.uni-heidelberg.de/$\sim$muon/lep/lep$_{-}$proc/K$_{-}$
Pachucki/kp.html).  

\bibitem{arno} R. Arnowitt, Phys. Rev. {\bf 92}, 1002 (1953);\\
W. A. Newcomb and E. E. Salpeter, Phys. Rev. {\bf 97}, 1146 (1955).

\bibitem{eide99} M. I. Eides,  H. Grotch and V. A. Shelyuto,
Phys. Rev. D {\bf 58}, 013008 (1998). 

\bibitem{newest} A. Czarnecki and K. Melnikov, 
Phys. Rev. Lett. {\bf 87}, 013001 (2001).

\bibitem{eide84} M. I. Eides and V. A. Shelyuto, Phys. Lett. {\bf 146} B,
241 (1984).

\bibitem{eide89} M. I. Eides, S. G. Karshenboim and V. A. Shelyuto, Phys.
Lett. B {\bf 216}, 405 (1989); S. G. Karshenboim, V. A. Shelyuto and M. I.
Eides, Yad. Fiz. {\bf 49}, 493 (1989)  (in Russian); Sov. J. Nucl. Phys.
{\bf 49}, 309 (1989).

\bibitem{pach96} K. Pachucki, Phys. Rev. A {\bf 54}, 1994 (1996).

\bibitem{nio} M. Nio and T. Kinoshita, Phys. Rev. D {\bf 55}, 7267 (1997).

\bibitem{blun} 
S. F. Blundell, K. T. Cheng and J. Sapirstein, Phys. Rev. Lett. 
{\bf 78}, 4914 (1997).


\bibitem{ys2001} V. A. Yerokhin and V. M. Shabaev, Phys. Rev. A {\bf 64}, 
012506 (2001).

\bibitem{kars99} S. G. Karshenboim, V. G. Ivanov and V. M.  Shabaev,
Phys. Sc. T{\bf 80} (1999) 491.

\bibitem{Nio2001}
M.~Nio, in {\em Quantum Electrodynamics and Physics of the Vacuum}. 
Ed. by G.~Cantatore, AIP, 2001, pp. 178--185.

\bibitem{sapi83a} J. R. Sapirstein, Phys. Rev. Lett. {\bf 51}, 985 (1983).

\bibitem{kars98} S. G. Karshenboim, V. G. Ivanov and V. M. Shabaev, 
Can. J. Phys. {\bf 76}, 503 (1998).


	
\bibitem{pach97} K. Pachucki, Phys.  Rev. Lett.  {\bf 79}, 4120 (1997).

\bibitem{sapi99} J. R. Sapirstein, private communication.

\bibitem{sapi83}  J. R. Sapirstein, E. A. Terray, D. R. Yennie, 
Phys.  Rev. Lett. {\bf 51}, 982 (1983).

\bibitem{eide86} M. I.  Eides,  S. G.  Karshenboim and V. A.  Shelyuto,
Phys. Lett. B {\bf 177}, 425 (1986).

\bibitem{broo} V. Yu. Brook, M. I. Eides, S. G.  Karshenboim  and  V. A.
Shelyuto, Phys. Lett. B {\bf 216}, 401 (1989).

\bibitem{eide88} M. I. Eides, S. G. Karshenboim  and  V. A.  Shelyuto,
Phys. Lett. B {\bf 202}, 572 (1988).


\bibitem{pach98} K. Pachucki and  S. G. Karshenboim, Phys. Rev. Lett.
{\bf 80}, 2101 (1998).

\bibitem{Czarnecki:1998zv}
A. Czarnecki, K. Melnikov, and A. Yelkhovsky,
Phys. Rev. Lett. {\bf 82}, 311 (1999).

\bibitem{Czarnecki:1999mw}
A. Czarnecki, K. Melnikov, and A. Yelkhovsky,
Phys. Rev. A {\bf 59}, 4316 (1999).

\bibitem{ej}
S.~Eidelman and F.~Jegerlehner, Z. Phys. C {\bf 67}, 585 (1995).

\bibitem{akh}
R.~R.~Akhmetshin {\it et al.}, Nucl. Phys. A {\bf 675}, 424c (2000).
 
\bibitem{ser}
S.~I.~Serednyakov, Nucl. Phys. B (Proc. Suppl.) {\bf 96}, 197 (2001). 

\bibitem{bes}
J.~Z.~Bai {\it et al.}, Phys. Rev. Lett. {\bf 84}, 594 (2000); \\
J.~Z.~Bai {\it et al.}, hep-ex/0102003, February 2001. 

\bibitem{pdg}
D.~E.~Groom {\it et al.}, Eur. Phys. J. {\bf C15}, 1  (2000).

\bibitem{kat}
S.~G.~Gorishny, A.~Kataev and S.~A.~Larin,  Phys. Lett. 
B {\bf 259}, 114 (1991);\\
L.~R.~Surguladze and M.~A.~Samuel, Phys. Rev. Lett. {\bf 66}, 560 (1991).

\bibitem{logash} 
I.~B.~Logashenko, Talk at the II Workshop on $e^+e^-$ Physics at 
Intermediate Energies, SLAC, April 30 -- May 2, 2001.

\bibitem{adh}
R.~Alemany, M.~Davier, and A.~H\"{o}cker,
Eur. Phys. J. C {\bf 2}, 123 (1998).

\bibitem{eid}
S.~I.~Eidelman, Nucl. Phys. B (Proc. Suppl.) {\bf 98}, 281 (2001).

\bibitem{kirt}
K.~Melnikov, SLAC-PUB-8844, hep-ph/0105267.
  
\bibitem{vena}
G.~Venanzoni, hep-ex/0106052.

\bibitem{bena}
M.~Benayoun, S.~I.~Eidelman, V.~N.~Ivanchenko, and Z.~K.~Silagadze, 
Mod. Phys. Lett. A {\bf 14}, 2605 (1999).

\bibitem{sol}
E.~P.~Solodov, Talk at the II Workshop on $e^+e^-$ Physics at 
Intermediate Energies, SLAC, April 30 -- May 2, 2001.

\bibitem{pich}
F.~Guerrero and A.~Pich, Phys. Lett. B {\bf 412}, 382 (1997).

\bibitem{barkov}
L.~M.~Barkov {\it et al.}, Nucl. Phys. B {\bf 256}, 365 (1985).

\bibitem{GeerIZ}
S.~Geer,
Phys.\ Rev.\ D {\bf 57}. 6989 (1998); D {\bf 59}, 039903 (E) (1998). 

\bibitem{AutinCI}
B.~Autin, A.~Blondel and J.~Ellis (eds.),
{\em Prospective study of muon storage rings at CERN},
Report CERN-99-02.

\bibitem{KunoWN}
Y.~Kuno,
Nucl.\ Instrum.\ Meth.\ A {\bf 451}, 233 (2000).

\bibitem{kniga}
{\em Hydrogen atom: Precision physics of simple atomic systems.} 
S. G. Karshenboim et al. (eds.), (Springer, Berlin, Heidelberg, 2001). 

\bibitem{kinoshita85} T. Kinoshita, B. Ni\v{z}i\'{c} and Y. Okamoto,
Phys. Rev. D {\bf 31}, 2108 (1985).

\end{thebibliography}
\end{document}